\documentclass[twocolumn,nofootinbib,showpacs,superscriptaddress,fleqn]{iopart}

\usepackage{amsfonts}
\usepackage{amstext}
\usepackage{amssymb}
\usepackage{bm}
\usepackage{cite}
\usepackage{dcolumn}
\usepackage{graphicx}
\usepackage{graphics}
\usepackage[latin1]{inputenc}
\usepackage{latexsym}
\usepackage{rotating}
\usepackage{url}
\usepackage[colorlinks=true]{hyperref}
\usepackage{xspace} 
\usepackage[usenames]{color}
\usepackage{booktabs}

\definecolor {darkgreen}{rgb}{0.2,0.7,0.2}

\newcommand\be{\begin{equation}}
\newcommand\ba{\begin{eqnarray}}
\newcommand\ee{\end{equation}}
\newcommand\ea{\end{eqnarray}}

\newcommand\bw{\begin{widetext}}
\newcommand\ew{\end{widetext}}

\newcommand{\nn}{\nonumber}

\newcommand{\K}{{\mbox{\tiny K}}}

\newcommand{\red}{{\mbox{\tiny red}}}

\newcommand{\sph}{{\mbox{\tiny sph}}}

\newcommand{\QG}{{\mbox{\tiny QG}}}
\newcommand{\EdGB}{{\mbox{\tiny EdGB}}}
\newcommand{\dCS}{{\mbox{\tiny dCS}}}
\newcommand{\scr}{{\mbox{\tiny scr}}}
\newcommand{\hor}{{\mbox{\tiny hor}}}

\newcommand{\MAT}{{\mbox{\tiny mat}}}

\begin{document}
\title[Black Hole Shadow as a Test of General Relativity]{Black Hole Shadow as a Test of General Relativity: Quadratic Gravity}

\author{%
Dimitry~Ayzenberg$^{1,2}$
and
Nicol\'as~Yunes$^{1}$
}

\address{$^{1}$~eXtreme Gravity Institute, Department of Physics, Montana State University, Bozeman, MT 59717, USA.}
\address{$^{2}$~Center for Field Theory and Particle Physics and Department of Physics, Fudan University, 200438 Shanghai, China}

\date{\today}

\begin{abstract} 

Observations of the black hole shadow of supermassive black holes, such as Sagittarius A* at the center of our Milky Way galaxy, allow us to study the properties of black holes and the nature of strong-field gravity. 
According to the Kerr hypothesis, isolated, stationary, and axisymmetric astrophysical black holes are described by the Kerr metric.
The Kerr hypothesis holds in General Relativity and in some modified gravity theories, but there are others in which it is violated. 
In principle, black hole shadow observations can be used to determine if the Kerr metric is the correct description for black holes, and in turn, they could be used to place constraints on modified gravity theories that do not admit the Kerr solution. 
We here investigate whether black hole shadow observations can constrain deviations from general relativity, focusing on two well-motivated modified quadratic gravity theories: Einstein-dilaton-Gauss-Bonnet gravity and dynamical Chern-Simons gravity. 
We show that current constraints on Einstein-dilaton-Gauss-Bonnet gravity are stronger than any that could be placed with black hole shadow observations of supermassive black holes. 
We also show that the same holds for dynamical Chern-Simons gravity through a systematic bias and a likelihood analysis when considering slowly-rotating supermassive black holes.  
However, observations of more rapidly-rotating black holes, with dimensionless spins $|\vec{J}|/M^{2}\simeq0.5$, could be used to better constrain dynamical Chern-Simons gravity.

\end{abstract}

\maketitle

\section{Introduction}

In April of 2017, the Event Horizon Telescope~\cite{2009astro2010S..68D} (EHT) undertook its first observing campaign that included (most of) the full worldwide radio telescope array. EHT, a very long baseline interferometer (VLBI), made observations of Sagittarius A*, the supermassive black hole (SMBH) at the center of our Milky Way galaxy. As the world's largest and most powerful VLBI, EHT has an observing resolution of about 23 $\mu$as, which is equivalent to standing in New York City and being able to read the date on a quarter in Los Angeles. Even with such a high resolution, resolving features on the scale of the event horizon of Sagittarius A* is quite difficult as its size, as seen from Earth, is only about 20 $\mu$as in diameter. Through complex data processing methods~\cite{0004-637X-850-2-172, Kuramochi:2018yea, 2017arXiv171101357B, 0004-637X-846-1-29, 0004-637X-838-1-1, 1538-3881-153-4-159, 0004-637X-832-2-156, 2016arXiv160900055B, Medeiros:2018dtc, 0004-637X-857-1-23}, these features can in principle be pulled out from the data and can provide a wealth of information about BHs, the environment very near them, and even gravity as a whole.

The primary feature of BHs that EHT is attempting to observe is the BH shadow, the dark region in electromagnetic observations caused by photons falling into the BH's event horizon. Whether a photon falls into the horizon or escapes to infinity is determined by the photon sphere, i.e.~the surface formed by all unstable spherical photon orbits around the BH. Any photon that crosses into the photon sphere will inexorably fall into the event horizon, creating a dark region in observations. The BH shadow, therefore, is a consequence of the strong-field gravity near the event horizon and it can in principle be used to determine the properties of the BH spacetime, such as its spin angular momentum. In fact, this is one of the main goals of EHT: to determine the spins of SMBHs, such as that of Sagittarius A* and the SMBH at the center of the M87 galaxy.

The BH shadow can also be used to test the \textit{Kerr hypothesis},~i.e.~that the correct description for all isolated, stationary, and axisymmetric astrophysical (uncharged) BHs is the Kerr metric. Under the Kerr hypothesis, any BH is completely determined by two parameters: the BH mass $M$ and (the magnitude of) the BH spin angular momentum $|\vec{J}|$. While the Kerr hypothesis holds in general relativity (GR) and in some modified gravity theories~\cite{Psaltis:2007cw}, there are a number of theories in which it does not~\cite{Alexander:2009tp}. Thus, in principle, observations of the BH shadow can also be used to test GR and place constraints on those modified gravity theories in which the Kerr hypothesis is violated.

A class of theories in which this is the case is quadratic gravity (QG)~\cite{Yunes:2011we}, and specifically, in two well-studied theories known as Einstein-dilaton-Gauss-Bonnet (EdGB) gravity~\cite{Moura:2006pz} and dynamical Chern-Simons (dCS) gravity~\cite{Alexander:2009tp}. In these theories, the Einstein-Hilbert action is modified through a dynamical scalar field that couples to a curvature invariant, the Gauss-Bonnet invariant in EdGB gravity and the Pontryagin invariant in dCS gravity. Both theories violate the strong equivalence principle, with dCS gravity also breaking parity invariance in the gravitational sector, two main pillars of GR~\cite{Will2014}. Black holes within EdGB gravity and dCS gravity have been studied extensively, and although a closed-form, exact solution for BHs with arbitrary spin has not yet been found, there is an ever-growing library of numerical and analytic solutions~\cite{Campbell:1990fu, Campbell:1990ai, Campbell:1991kz, Campbell:1992hc, Alexeev:1996vs, Kanti:1995vq, Torii:1996yi, Kleihaus:2011tg, PhysRevD.83.104002, Yunes:2009hc, Pani:2011gy, PhysRevD.86.044037, Ayzenberg:2014aka, McNees:2015srl}. In this paper, we will use a pair of closed-form, analytic, yet approximate solutions in EdGB and in dCS gravity, which represent slowly-rotating BHs up to quintic order in the ratio of the BH spin to its mass squared~\cite{Maselli:2017kic, Maselli:2015tta}. 

The primary goal of this paper is to determine whether constraints can be placed on EdGB gravity and on dCS gravity using observations of BH shadows. The current constraints on either theory are not very stringent as they come from observations that are not in the strong-field gravity regime~\cite{kent-LMXB, alihaimoud, PhysRevD.86.044037}. BH shadows can in principle probe this regime, and so they may lead to stronger constraints, assuming the observational error and modeling systematics are under control and do not dominate over any possible modified gravity effects. As the data of the EHT observing campaign are still being analyzed, it is difficult to estimate the expected observational and systematic error; a recent study has estimated that EHT will have an error of roughly $10\%$ on the extracted observables~\cite{Psaltis:2014mca}, and we will use this estimate in our work.

In order to achieve this goal, we carry out a couple of data analysis studies on simulated BH shadow observations with modified gravity theory models. In both cases, we simulate the shadow of a Kerr BH and treat it as the \textit{observation} or \textit{injection}, while we simulate shadows of BHs in modified gravity and treat them as the \textit{model} we fit to the injection. The model, thus, depends on a GR deformation parameter (a certain combination of the coupling constants in EdGB and dCS gravity), the BH spin, and the inclination angle, while the injection depends only on the injected spin and the injected inclination angle.  

In the first study, we fit for the value of the BH spin by minimizing the $\chi^{2}$ between the injection and the model, for a fixed value of the deformation parameter and the inclination angle. In principle, since the model is different from the injection, we expect the best-fit spin parameter to be biased away from the injected value, with the systematic error growing as we fix larger values of the deformation parameter. When the systematic error becomes larger than the expected statistical error on the extraction of the spin, then we can infer that the deformation parameter may be measurable at that level.       

The second study is a likelihood analysis. We calculate the $\chi^{2}$ statistic between the GR injection and the modified gravity model, varying over \emph{all} parameters of the model and using flat priors on all parameters, to construct the likelihood function. We then marginalize the likelihood over the spin parameter and the inclination angle to construct the marginalized posterior probability distribution for the deformation parameter. If the posterior peaks around zero deformation, we can then infer the accuracy to which the deformation parameter could be constrained. If instead the posterior is similar to the prior, we can then infer that the injection cannot distinguish between a GR and an EdGB or dCS model. 

The main result of our paper is that EdGB gravity cannot by constrained using BH shadow observations, while it may be possible to constrain dCS gravity with observations of BHs, provided the observed dimensionless spin $\chi\gtrsim0.5$. In the case of EdGB gravity, the current constraints, coupled to the large masses of the SMBHs that are targets for BH shadow observations, leads to a GR deformation that is over 20 orders of magnitude smaller than the GR contribution. Thus, it is clear that SMBH shadow observations cannot be used to constrain EdGB gravity. 

In the case of dCS gravity, the GR deformation is too small to be observable for slowly-rotating BHs. We find no bias in the recovered BH spin parameter and the marginalized posterior on the deformation parameter is very similar to the flat prior for injected spins in the range $|\vec{J}|/M^{2}\lesssim0.45$. However, BH shadow observations of more rapidly-rotating BHs, with $|\vec{J}|/M^{2}\gtrsim 0.5$, could constrain dCS gravity. While we find no significant bias in the recovered spin parameter even when $|\vec{J}|/M^{2} = 0.5$, we do find that the the goodness of fit deteriorates with increasing spin. We also find that the marginalized posterior on the deformation parameter does become peaked at zero and significantly different from the prior for large injected spin parameter. These conclusions could be verified in the future by repeating our analysis with a numerical dCS BH metric valid for moderate or large spins, such as the rapidly-rotating solution of~\cite{McNees:2015srl}.

The remainder of this paper presents the details of the calculations pertaining to these results. Section~\ref{sec:QG} summarizes QG, discusses the BH solutions used in this paper, and makes the argument for our conclusion on EdGB gravity. Section~\ref{sec:shadow} describes the BH shadow and how it is calculated numerically. Section~\ref{sec:shad-QG} details the methodology and results of the systematic bias study and the likelihood analysis. Section~\ref{sec:disc} concludes by summarizing our results and discussing implications. Throughout, we use the following conventions: the metric signature $(-,+,+,+)$; Latin letters in index lists stand for spacetime indices; parentheses and brackets in index lists for symmetrization and antisymmetrization, respectively, i.e. $A_{(ab)}=(A_{ab}-A_{ba})/2$ and $A_{[ab]}=(A_{ab}-A_{ba})/2$; geometric units with $G=c=1$ (e.g. $1 M_\odot$ becomes 1.477 km by multiplying by $G/c^2$ or $4.93\times10^{-6}$ s by multiplying by $G/c^3$), except where otherwise noted.

\section{Quadratic Gravity and BH Solutions}
\label{sec:QG}

 In QG the Einstein-Hilbert action is modified by including all possible quadratic, algebraic curvature scalars with running (i.e.~nonconstant) couplings~\cite{Yunes:2011we}
 \begin{eqnarray}
 S&\equiv\int d^4x\sqrt{-g}\{\kappa R+\alpha_1f_1(\vartheta)R^2+\alpha_2f_2(\vartheta)R_{ab}R^{ab}
\nonumber \\
&+\alpha_3f_3(\vartheta)R_{abcd}R^{abcd}+\alpha_4f_4(\vartheta)R_{abcd}\text{}^{*}R^{abcd}
\nonumber \\
&-\frac{\beta_{\QG}}{2}\left[\nabla_a\vartheta\nabla^a\vartheta+2V(\vartheta)\right]+\mathcal{L}\MAT\}.
 \end{eqnarray}
 Here, $g$ is the determinant of the metric $g_{ab}$, $R_{ab}$, $R_{abcd}$, and $\text{}^{*}R_{abcd}$ are the Ricci scalar, Ricci tensor, and the Riemann tensor and its dual, respectively, with the latter defined as
 \begin{equation}
 \text{}^{*}R^a_{~bcd}= \frac{1}{2} \varepsilon_{cd}^{~~ef}R^a_{~bef}\,,
 \end{equation}
 and $\varepsilon^{abcd}$ is the Levi-Civita tensor. $\mathcal{L}\MAT$ is the external matter Lagrangian, $\vartheta$ is a field, $f_{i}(\vartheta)$ are functionals of this field, $(\alpha_{i},\beta_{\QG})$ are coupling constants, and $\kappa=1/(16\pi)$.
 
Two theories of particular interest within QG are EdGB gravity and dCS gravity. In EdGB gravity, $(\alpha_{1},\alpha_{2},\alpha_{3},\alpha_{4})=(\alpha_{\EdGB},-4\alpha_{\EdGB},\alpha_{\EdGB},0)$ and $(f_{1},f_{2},f_{3},f_{4})=(e^{\vartheta},e^{\vartheta},e^{\vartheta},0)$, where $\alpha_{\EdGB}$ is the EdGB gravity coupling constant and $\vartheta$ is the dilaton. In dCS gravity, $(\alpha_{1},\alpha_{2},\alpha_{3},\alpha_{4})=(0,0,0,\alpha_{\dCS}/4)$ and $(f_{1},f_{2},f_{3},f_{4})=(0,0,0,\vartheta)$, where $\alpha_{\dCS}$ is the dCS coupling constant and $\vartheta$ is the dCS (axion-like) field. The strongest constraint on EdGB gravity comes from low-mass X-ray binary observations, $\sqrt{|\alpha_{\EdGB}|}<1.9\times10^{5}$cm~\cite{kent-LMXB}. The strongest constraint on dCS gravity comes from Solar System~\cite{alihaimoud} and tabletop experiments~\cite{PhysRevD.86.044037}, $\sqrt{|\alpha_{\dCS}|}<10^{13}$cm.
 
We now introduce a dimensionless coupling parameter
 \begin{equation}
 \zeta=\frac{\alpha^{2}}{\kappa M^{4}},
 \end{equation}
 where $M$ is the typical mass of the system, in our case the BH mass, $\alpha$ is either $\alpha_{\EdGB}$ or $\alpha_{\dCS}$ for EdGB gravity and dCS gravity respectively, and we have set $\beta_{\QG}=1$. The primary targets for BH shadow observations by telescopes such as EHT are SMBHs, which have masses in the range $10^{6}M_{\odot}\lesssim M \lesssim 10^{10}M_{\odot}$. Using these SMBH masses and the current constraints on EdGB gravity and dCS gravity we can calculate the maximum value $\zeta$ can take in each case for SMBH shadow observations. For EdGB gravity $\zeta_{\EdGB}\lesssim 10^{-22}$ and for dCS gravity $\zeta_{\dCS}\lesssim 10^{9}$ in shadow observations; this last bound is superseded by the small coupling approximation $\zeta_{\dCS} \ll 1$, which is required in dCS gravity as it is an effective gravity theory~\cite{Alexander:2009tp}. 
 
Since the dimensionless parameter $\zeta$ controls the magnitude of the GR deviation, we can already conclude that BH shadow observations will not be able to constrain EdGB gravity:~given the masses of SMBHs and the current constraints on EdGB gravity, the maximum deviation from BH shadows in GR expected in this theory is too small to be detected with current and planned BH shadow observations. Perhaps BH shadow observations of stellar-mass BHs would have measurable deviations and could be used to place constraints on EdGB, but such observations are unlikely in the near future. For this reason, we will focus on dCS gravity for the remainder of this paper, which is not as well constrained currently.
 
Let us now discuss BH solutions in GR and in dCS gravity. In GR, the solution for an isolated, stationary, axisymmetric, and uncharged BH is the Kerr metric. The line element associated with this metric in Boyer-Lindquist (BL) coordinates ($t,r,\theta,\phi$) is given by
\begin{eqnarray}
ds_{\K}^2=&-\left(1-\frac{2Mr}{\Sigma_{\K}}\right)dt^2-\frac{4Mar\sin^2\theta}{\Sigma_{\K}}dtd\phi+\frac{\Sigma_{\K}}{\Delta_{\K}}dr^2
\nonumber \\
&+\Sigma_{\K} 
d\theta^2+\left(r^2+a^2+\frac{2Ma^2r\sin^2\theta}{\Sigma_{\K}}\right)\sin^2\theta 
d\phi^2,
\label{eq:Kerr-metric}
\end{eqnarray}
with $\Delta_{\K}\equiv r^2-2Mr+a^2$ and $\Sigma_{\K}\equiv r^2+a^2\cos^2\theta$. Here $M$ is the mass of the BH and $a\equiv J/M$ is the Kerr spin parameter, where $J:=|\vec{J}|$ is the magnitude of the BH spin angular momentum.

In dCS gravity, we will focus on the approximate, stationary, and axisymmetric solution that represents a slowly-rotating BH to fifth order in the spin~\cite{Maselli:2017kic}. In Boyer-Lindquist like coordinates, this solution takes the form
\begin{equation}
g_{ab}^{\dCS}=g_{ab}^{\K}+\zeta_{\dCS} \sum_{i=0}^{5} \chi^{i} g^{[i]}_{ab}, \label{eq:dCS}
\end{equation}
where $g_{ab}^{\K}$ is the Kerr metric in Boyer-Lindquist coordinates, $\chi=a/M=|\vec{J}|/M^{2}$ is the dimensionless spin parameter, and $g^{[i]}_{ab}$ are metric deformations due to dCS gravity, with $[i]$ representing the order in the spin~\cite{Cardenas-Avendano:2018ocb}. 

A particularly important property of the BH solutions for calculating the BH shadow is the location of the event horizon. The horizon is defined as a null surface formed by marginally-trapped, null geodesics. The normal to the surface $n^{a}$ must be null, and thus the event horizon satisfies the horizon equation
\begin{equation}
g^{ab}\partial_{a}F\partial_{b}F=0, \label{eq:hor}
\end{equation}
where $F(x^{a})$ is a level surface function with normal $n_{a}=\partial_{a}F$. Both the Kerr and dCS BH spacetimes are stationary, axisymmetric, and reflection symmetric about the poles and equator, and thus the level surfaces are only dependent on radius $r$. Then, without loss of generality, we can let $F(x^{a})=r-r_{\hor}$, where $F=0$ defines the location of the horizon. Equation~(\ref{eq:hor}) becomes $g^{rr}=0$, and solving this equation for the Kerr metric one finds
\begin{equation}
 r^{\K}_{\hor}=M(1+\sqrt{1-\chi^{2}}).
 \end{equation}

In dCS gravity, however, the location of the horizon should not be at the Kerr location, a fact that is obscured by the slow-rotation approximation inherent in Eq.~(\ref{eq:dCS}). In fact, if one treats the dCS BH solution as \textit{exact},~i.e.~without further expanding in $\chi$ or $\zeta_{\dCS}$, one would mistakenly conclude that the horizon is at $r^{\dCS}_{\hor}=2M$. This mistake would recur even if one studies the slow-rotation expansion of the Kerr metric, and it can lead to unphysical behavior,~e.g.~one could conclude that there are no photon orbits outside of the horizon (see~\cite{Ayzenberg:2016ynm} for a discussion of such behavior in the slow-rotation expansion of the Kerr metric). As we will explain in the next section, in this paper we must treat the solution as exact and so such unphysical behavior must be dealt with prior to any analysis.

In general, such unphysical behavior and other spurious features can be eliminated by performing a resummation,~i.e.~introducing terms to the metric that are higher order in $\chi$ and have not been explicitly calculated but ought to appear at higher order and/or in the exact solution. Making the correct choice of resummation, though, can be quite difficult as there are in principle an infinite number of ways to resum the metric, and the exact solution of a rotating BH in dCS gravity is unknown. For our purposes, it suffices to resum the metric of Eq.~(\ref{eq:dCS}) such that the event horizon is at the location it would be if one solved Eq.~(\ref{eq:hor}) perturbatively to first order in $\zeta_{\dCS}$ and $5$th order in the spin $\chi$~\cite{Yagi:2012ya}:
\begin{equation}
r^{\dCS}_{\hor}=2M\left(1-\frac{1}{4}\chi^{2}-\frac{1}{16}\chi^{4}\right)-\frac{915}{28672}\zeta_{\dCS} M\chi^{2}\left(1+\frac{351479}{439200}\chi^{2}\right).
\end{equation}

One choice of resummation that accomplishes this is to replace any appearances in the solution of $f(r)=1-2M/r$ with $f_{\dCS}(r)=1-r^{\dCS}_{\hor}/r$ and any appearances of $\Delta_{\K}$ with $\Delta_{\dCS}$ such that when solving $\Delta_{\dCS}=0$ for $r$ and expanding to $\mathcal{O}(\zeta_{\dCS},\chi^{5})$ one finds that $r=r^{\dCS}_{\hor}$. $\Delta_{\dCS}$ is given by
\begin{equation}
\Delta_{\dCS}=\Delta_{\K}-\frac{915}{14336}\zeta_{\dCS} a^{2}\left(1+\frac{351479}{439200}\frac{a^{2}}{M^{2}}\right).
\end{equation}
To retain the asymptotic behavior in the $\chi<<1$ limit, we add the following counterterm to the $(r,r)$ component of the metric
{\setlength{\mathindent}{0pt}
\begin{eqnarray}
\delta g_{rr}&=\frac{915}{14336}\zeta_{\dCS}\chi^{2}\frac{M^{2}}{r^{2}f^{2}_{\dCS}}+\frac{2717}{14336}\zeta_{\dCS}\chi^{4}\frac{M^{4}}{r^{4}f^{3}_{\dCS}}\left[\left(1+\frac{402}{2717}\frac{M}{r}-\frac{1704}{19019}\frac{M^{2}}{r^{2}}+\frac{458960}{19019}\frac{M^{3}}{r^{3}}\right.\right.
\nonumber \\
&\left.\left.-\frac{1874640}{19019}\frac{M^{4}}{r^{4}}-\frac{6768}{209}\frac{M^{5}}{r^{5}}-\frac{911712}{2717}\frac{M^{6}}{r^{6}}+\frac{2903040}{2717}\frac{M^{7}}{r^{7}}\right)\left(3\cos^{2}\theta-1\right)\right.
\nonumber \\
&\left.+\frac{631}{6240}\frac{r^{2}}{M^{2}}\left(1+\frac{175442}{131879}\frac{M}{r}-\frac{732000}{131879}\frac{M^{2}}{r^{2}}-\frac{785600}{131879}\frac{M^{3}}{r^{3}}-\frac{75008}{6941}\frac{M^{4}}{r^{4}}\right.\right.
\nonumber \\
&\left.\left.-\frac{126195776}{395637}\frac{M^{5}}{r^{5}}-\frac{593814080}{2769459}\frac{M^{6}}{r^{6}}+\frac{511400640}{923153}\frac{M^{7}}{r^{7}}+\frac{10560693760}{2769459}\frac{M^{8}}{r^{8}}\right.\right.
\nonumber \\
&\left.\left.+\frac{134859520}{20823}\frac{M^{9}}{r^{9}}-\frac{25506929920}{923153}\frac{M^{10}}{r^{10}}+\frac{28366106880}{923153}\frac{M^{11}}{r^{11}}-\frac{64275077120}{923153}\frac{M^{12}}{r^{12}}\right.\right.
\nonumber \\
&\left.\left.+\frac{8959749120}{131879}\frac{M^{13}}{r^{13}}+\frac{5270372352}{131879}\frac{M^{14}}{r^{14}}\right)\right].
\end{eqnarray}}
For completeness the resummed dCS metric is given in~\ref{app:dCSsols}. We use this metric throughout the rest of this work.
 
\section{Black Hole Shadow}
\label{sec:shadow}

The BH shadow is the observable consequence of the photon sphere of a BH spacetime. The later is the surface formed from the combination of all unstable and spherical photon orbits,~i.e.~the separatrix between photon geodesics that escape to spatial infinity and those that fall into the event horizon~\cite{Claudel:2000yi}. For generic spacetimes, the null-geodesic equations that describes the photon geodesics cannot be solved for analytically and the shadow must be calculated by solving for the photon motion numerically. For separable spacetimes, however, such as the Kerr metric in BL coordinates, the photon sphere has an analytic solution and the shadow can be calculated from that solution. We explicitly present the calculation of the photon sphere and shadow for the Kerr metric in BL coordinates in~\ref{app:shadow}. The BH metric in dCS gravity studied here is not separable and the shadow must be solved for numerically. We do so by using a general relativistic ray-tracing code described here.

Our ray-tracing code computes the trajectories of photons near the BH following the method described in~\cite{Psaltis:2010ww}. In all stationary and axisymmetric spacetimes, the specific energy $E$ and the $z$-component of the specific angular momentum $L_{z}$ are conserved quantities and related to components of the four-momentum of a test particle via $p_{t}=-E$ and $p_{\phi}=L_{z}$. These relations can be rewritten to get two first-order differential equations for the evolution of the $t$- and $\phi$-components of the photon position
\begin{eqnarray}
\frac{dt}{d\lambda'}&=-\frac{bg_{t\phi}+g_{\phi\phi}}{g_{tt}g_{\phi\phi}-g_{t\phi}^{2}}, \label{eq:dt}
\\
\frac{d\phi}{d\lambda'}&=b\frac{g_{tt}+g_{t\phi}}{g_{tt}g_{\phi\phi}-g_{t\phi}^{2}}, \label{eq:dphi}
\end{eqnarray}
where $\lambda'\equiv E/\lambda$ is the normalized affine parameter and $b\equiv L_{z}/E$ is the impact parameter.

For the $r$- and $\theta$-components of the photon position, we solve the second-order geodesic equations for a generic axisymmetric metric
\begin{eqnarray}
\frac{d^{2}r}{d\lambda'^{2}}=&-\Gamma^{r}_{tt}\left(\frac{dt}{d\lambda'}\right)^{2}-\Gamma^{r}_{rr}\left(\frac{dr}{d\lambda'}\right)^{2}-\Gamma^{r}_{\theta\theta}\left(\frac{d\theta}{d\lambda'}\right)^{2}-\Gamma^{r}_{\phi\phi}\left(\frac{d\phi}{d\lambda'}\right)^{2}
\nonumber \\
&-2\Gamma^{r}_{t\phi}\left(\frac{dt}{d\lambda'}\right)\left(\frac{d\phi}{d\lambda'}\right)-2\Gamma^{r}_{r\theta}\left(\frac{dr}{d\lambda'}\right)\left(\frac{d\theta}{d\lambda'}\right),\label{eq:d2r}
\\ \nonumber \\
\frac{d^{2}\theta}{d\lambda'^{2}}=&-\Gamma^{\theta}_{tt}\left(\frac{dt}{d\lambda'}\right)^{2}-\Gamma^{\theta}_{rr}\left(\frac{dr}{d\lambda'}\right)^{2}-\Gamma^{\theta}_{\theta\theta}\left(\frac{d\theta}{d\lambda'}\right)^{2}-\Gamma^{\theta}_{\phi\phi}\left(\frac{d\phi}{d\lambda'}\right)^{2}
\nonumber \\
&-2\Gamma^{\theta}_{t\phi}\left(\frac{dt}{d\lambda'}\right)\left(\frac{d\phi}{d\lambda'}\right)-2\Gamma^{\theta}_{r\theta}\left(\frac{dr}{d\lambda'}\right)\left(\frac{d\theta}{d\lambda'}\right),\label{eq:d2theta}
\end{eqnarray}
where $\Gamma^{a}_{bc}$ are the Christoffel symbols of the metric. Solving this second-order system renders our ray-tracing code applicable to non-separable spacetimes.

\begin{figure}
\includegraphics[width=\columnwidth,trim={7cm 7cm 7cm 7cm},clip]{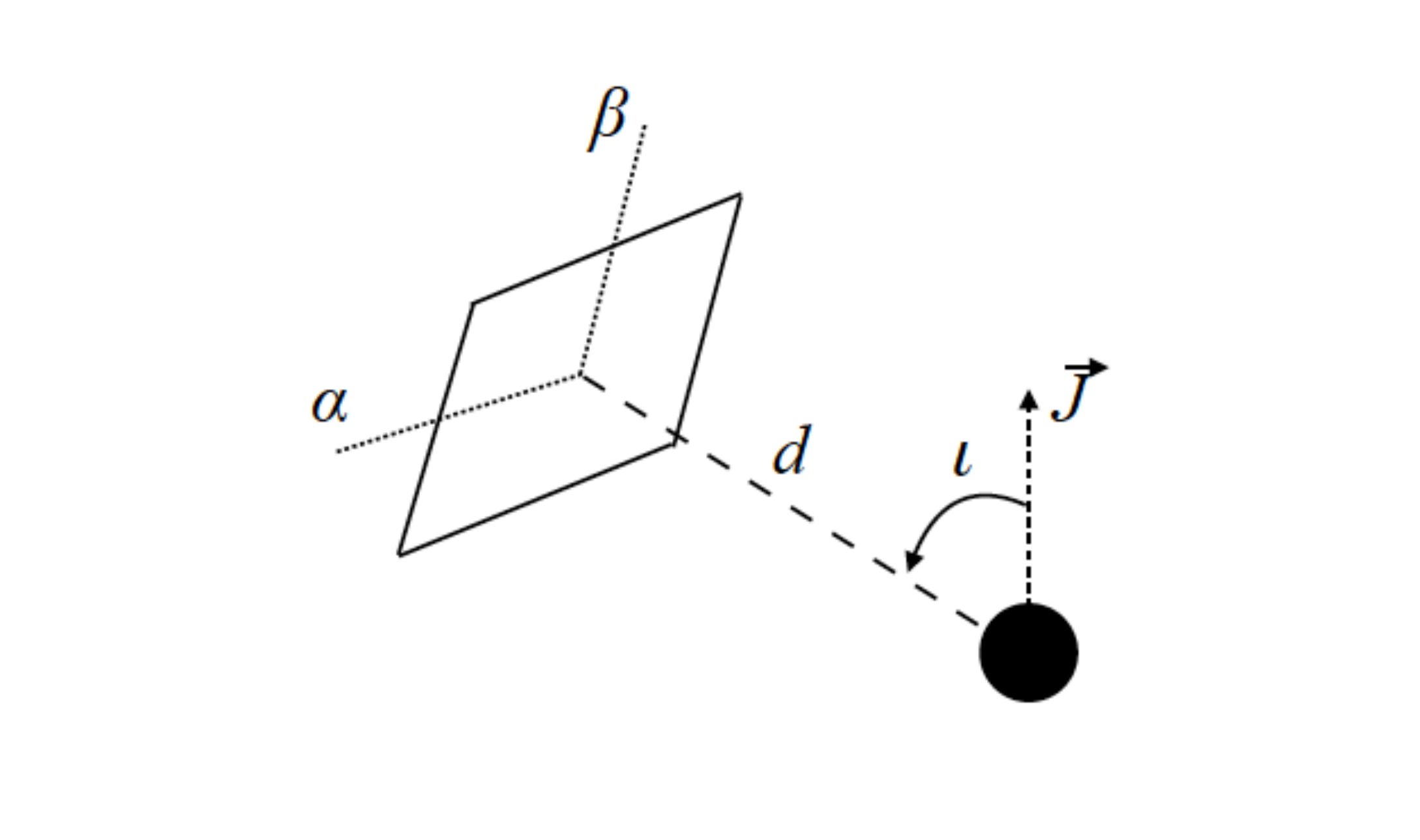}
\caption{\label{fig:orient} Depiction of the orientation of the observing screen relative to the BH. $\alpha$ and $\beta$ are the celestial coordinates on the observer's sky, $d$ is the distance to the BH, $\vec{J}$ is the spin angular momentum of the BH, and $\iota$ is the inclination angle between $\vec{J}$ and the observer's line of sight.}
\end{figure}

The reference frame and coordinate system is chosen such that the BH is stationary at the origin and the BH's spin angular momentum is along the $z$-axis. In the code and for the remainder of this paper, we use units with the BH mass $M=1$, since the shape of the BH shadow is independent of the BH mass and only the size of the shadow changes with mass. For the numerical evolution, the observing screen is centered at a distance $d=1000$, the azimuthal angle $\theta=\iota$, and the polar angle $\phi=0$. On the screen, we use polar coordinates $r_{\text{scr}}$ and $\phi_{\scr}$, which are related to the celestial coordinates $(\alpha,\beta)$ on the observer's sky by $\alpha=r_{\scr}\cos\phi_{\scr}$ and $\beta=r_{\scr}\sin\phi_{\scr}$. Figure~\ref{fig:orient} depicts the orientation of the screen and BH.

As we know the final positions and momenta of the photons, but not where they originated from, we evolve our system of equations (Eqs.~\ref{eq:dt}-\ref{eq:d2theta}) backwards in time. We initialize each photon with some initial position on the screen and a four-momentum that is perpendicular to the screen. The latter condition simulates placing the screen very far away from the observer, as only those photons that are moving perpendicular to the screen at a distance $d$ will also hit a screen at spatial infinity.

The initial position and four-momentum of each photon in the BL coordinates of the BH spacetime is given by
\begin{eqnarray}
r_{i}=&\left(d^{2}+\alpha^{2}+\beta^{2}\right)^{1/2},
\\
\theta_{i}=&\arccos\left(\frac{d\cos\iota+\beta\sin\iota}{r_{i}}\right),
\\
\phi_{i}=&\arctan\left(\frac{\alpha}{d\sin\iota-\beta\cos\iota}\right),
\end{eqnarray}
and
\begin{eqnarray}
\left(\frac{dr}{d\lambda'}\right)_{i}&=\frac{d}{r_{i}},
\\
\left(\frac{d\theta}{d\lambda'}\right)_{i}&=\frac{-\cos\iota+\frac{d}{r_{i}^{2}}\left(d\cos\iota+\beta\sin\iota\right)}{\sqrt{r_{i}^{2}-\left(d\cos\iota+\beta\sin\iota\right)^{2}}},
\\
\left(\frac{d\phi}{d\lambda'}\right)_{i}&=\frac{-\alpha\sin\iota}{\alpha^{2}+\left(d\sin\iota-\beta\cos\iota\right)^{2}},
\\
\left(\frac{dt}{d\lambda'}\right)_{i}&=-\frac{g_{t\phi}}{g_{tt}}\left(\frac{d\phi}{d\lambda'}\right)_{i}-\left[\frac{g_{t\phi}^{2}}{g_{tt}^{2}}\left(\frac{d\phi}{d\lambda'}\right)_{i}^{2}\right.
\nonumber \\
&\left.-\left(g_{rr}\left(\frac{dr}{d\lambda'}\right)_{i}^{2}+g_{\theta\theta}\left(\frac{d\theta}{d\lambda'}\right)_{i}^{2}+g_{\phi\phi}\left(\frac{d\phi}{d\lambda'}\right)_{i}^{2}\right)\right]^{1/2}.
\end{eqnarray}
The component $\left(dt/d\lambda'\right)_{i}$ is found by requiring the norm of the photon four-momentum to be zero. The conserved quantity $b$ is computed from the initial conditions, as this quantity is required in Eqs.~(\ref{eq:dt}) and~(\ref{eq:dphi}).

Instead of finely sampling initial conditions over the entire screen, we speed up the code as follows. For each value of $\phi_{\scr}$ in the range $0\leq\phi_{\scr}\leq\pi$ and with steps of $\pi/180$, we search inside $0\leq r_{\scr}\leq10$ for the location of the boundary of the BH shadow. This boundary is the separatrix between photons that fall into the horizon and photons that go out to spatial infinity. The former are determined by any photons crossing $r=r_{\hor}+\delta r$ with $\delta r=10^{-6}$, while the latter are determined by photons that reach $r>d=1000$. We then zoom in on the boundary to an accuracy of $\delta r_{\scr}\sim 10^{-3}$ to accurately determine the value of $r_{\scr}$ that corresponds to the shadow boundary for the current value of $\phi_{\scr}$. Using this methodology, we can accurately calculate the BH shadow much more efficiently than if the entire screen were finely sampled. The drawback of this method is that secondary shadow features besides the boundary can be missed; we have checked that the shadow for a dCS BH does not contain any secondary features.

\section{Shadows of dCS BHs}
\label{sec:shad-QG}

In this section we describe our methodology for analyzing and comparing the BH shadow of dCS gravity to the shadow of a Kerr BH, present the results of our analysis, and discuss whether it is possible for observations of the shadow of BHs to test GR and place better-than-current constraints on dCS gravity.

\subsection{Characterization of the BH Shadow}
\label{subsec:characterization}

\begin{figure}
\centering
\includegraphics[trim={0cm 0cm 5cm 0cm}]{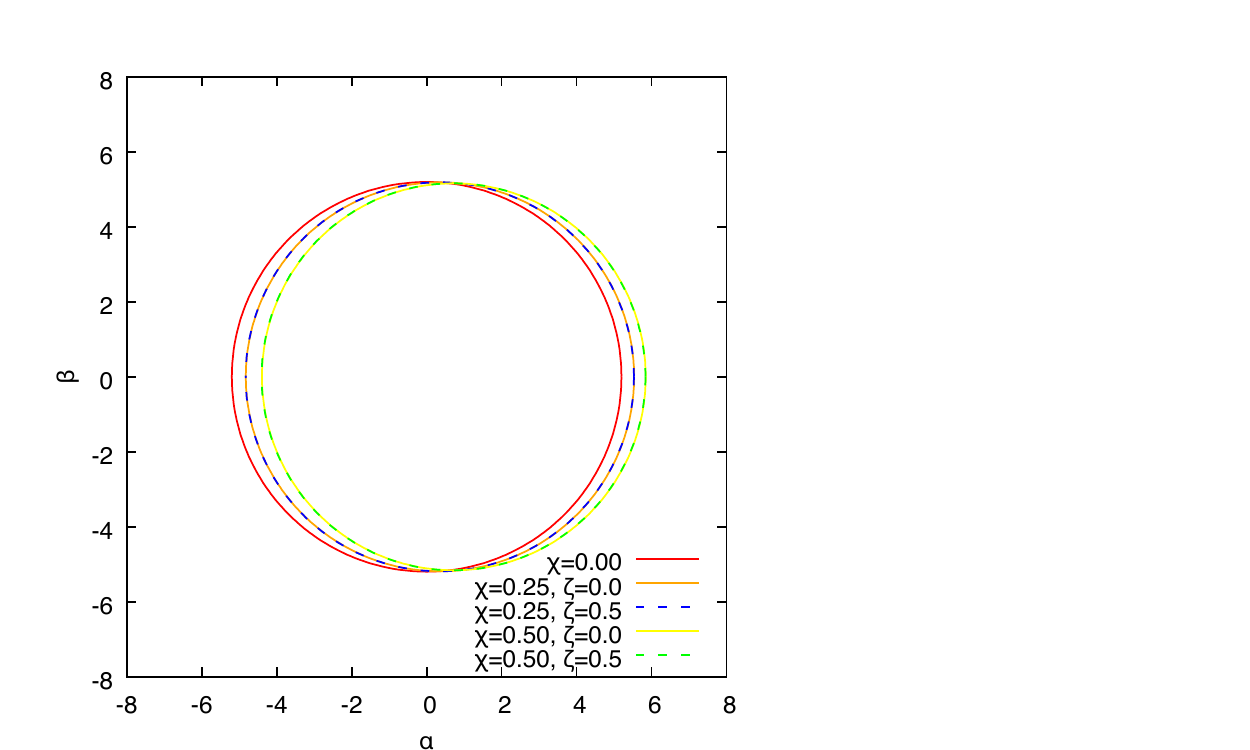}
\caption{\label{fig:shadows} Black hole shadows of the Kerr BH (solid lines) and dCS gravity BH (dashed lines) for $\chi=(0,0.25,0.5)$, $\iota=\pi/4$, and $\zeta_{\dCS}=(0,0.5)$.}
\end{figure}

We create a synthetic BH shadow observation using the Kerr metric and try to fit this shadow with a dCS shadow model, using the ray-tracing code described in Sec.~\ref{sec:shadow}. The dCS shadow model depends on the spin parameter $\chi$, the inclination angle $\iota$, and the the coupling constant or deformation parameter $\zeta_{\dCS}$. We generate models throughout this 3-dimensional parameter space in the prior ranges $0\leq\chi\leq0.5$, $0\leq\iota\leq\pi/2$, and $0\leq\zeta_{\dCS}\leq0.5$ and $0 \leq \iota \leq \pi/2$. The priors on $\chi$ and $\zeta_{\dCS}$ are to remain consistent with the slow-rotation and small-coupling approximations required in the dCS gravity BH solution. Figure~\ref{fig:shadows} shows the shadows for $\chi=(0,0.25,0.5)$, $\iota=\pi/4$, and $\zeta_{\dCS}=(0,0.5)$. Observe that the dCS gravity shadows are not noticeably different from the Kerr shadows for the same values of spin.

For the purposes of our analysis, and to more easily see the differences between the shadows in the Kerr solution and the dCS gravity solution, we parameterize the shape of the shadow in terms of the horizontal displacement from the center of the image $D$, the average radius of the sphere $\langle R\rangle$, and the asymmetry parameter $A$. There are other ways to model the shape of the shadow (see~e.g.~\cite{Tsukamoto:2014tja, Abdujabbarov:2015xqa}), but the results of this work should be similar regardless of the chosen parameterization. The horizontal displacement $D$ is the shift of the center of the shadow from the center of the BH, and it is defined by
\begin{equation}
D\equiv\frac{|\alpha_{\text{max}}-\alpha_{\text{min}}|}{2},
\end{equation}
where $\alpha_{\text{min}}$ and $\alpha_{\text{max}}$ are the minimum and maximum horizontal coordinates of the image on the observing screen, respectively. As the Kerr spacetime and the dCS gravity BH spacetime studied in this work are axially symmetric, there is no vertical displacement of the image. The average radius $\langle R\rangle$ is the average distance of the boundary of the shadow from its center, and it is defined by
\begin{equation}
\langle R\rangle\equiv\frac{1}{2\pi}\int_{0}^{2\pi}R(\vartheta)d\vartheta,
\end{equation}
where $R(\vartheta)\equiv\left[(\alpha-D)^{2}+\beta(\alpha)^{2}\right]^{1/2}$ and $\vartheta\equiv\tan^{-1}[\beta(\alpha)/\alpha]$. The asymmetry parameter is the distortion of the shadow from a circle, and it is defined by
\begin{equation}
A\equiv2\left[\frac{1}{2\pi}\int_{0}^{2\pi}(R-\langle R\rangle)^{2}d\vartheta\right]^{1/2}.
\end{equation}

\begin{figure}
\centering
\includegraphics[]{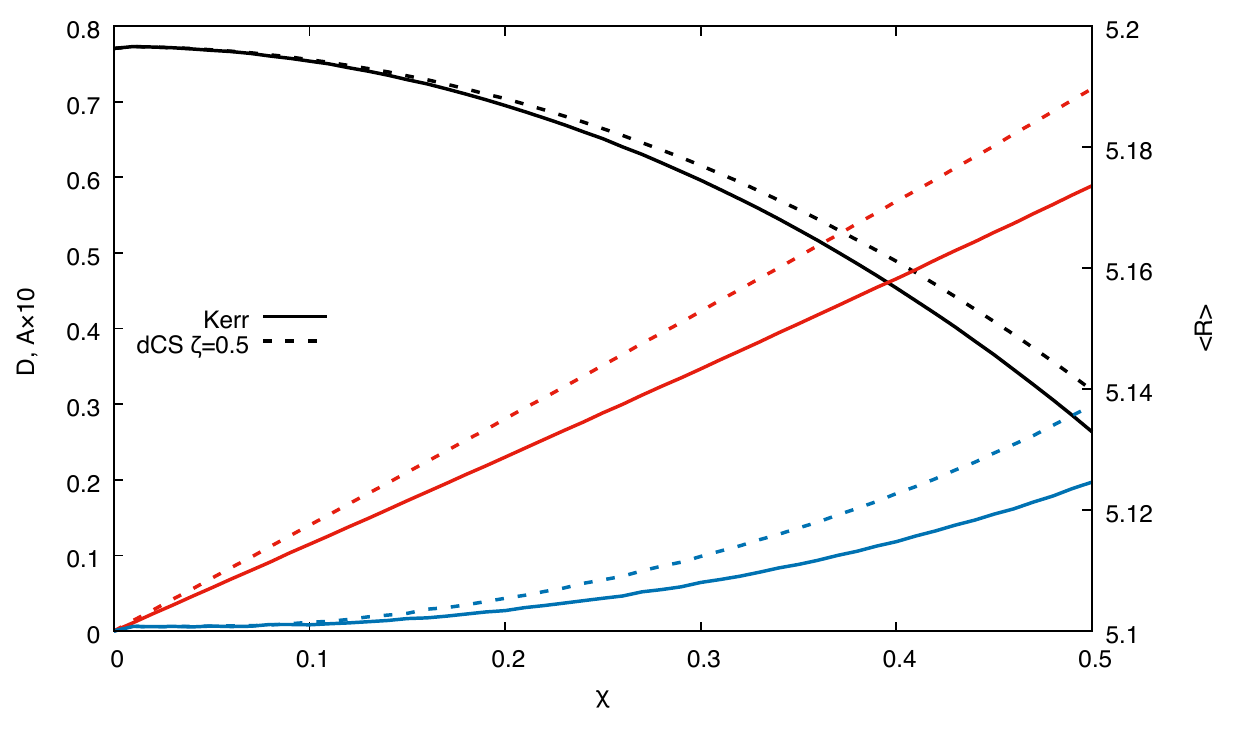}
\caption{\label{fig:params} Displacement $D$ (red), average radius $\langle R\rangle$ (black), and asymmetry parameter $A$ (blue) for BH shadows of the Kerr solution (solid) and the dCS gravity BH solution with $\zeta_{\dCS}=0.5$ (dashed) at an inclination angle $\iota=\pi/4$.}
\end{figure}

Figure~\ref{fig:params} shows $D$, $\langle R\rangle$, and $A$, as a function of the spin parameter $\chi$ for fixed values of the coupling parameter $\zeta_{\dCS}=(0,0.5)$ at an inclination angle of $\iota=\pi/4$. The average radius $\langle R\rangle$ is not significantly different between Kerr and dCS gravity, but at a spin of $\chi=0.5$ the displacement $D$ has about a $20\%$ difference and the asymmetry parameter $A$ has about a $50\%$ difference. Whether those differences are enough to place a constraint on dCS gravity with BH shadow observations requires a more precise analysis, as we do in the following subsection.

%

\subsection{Projected Constraints}

Let us now study whether constraints can be placed on dCS gravity using BH shadow observations. We follow the same prescription of~\cite{Ayzenberg:2016ynm} and assume that future observations are of Kerr shadows. We refer to the Kerr shadow as the \textit{injected synthetic signal} or \textit{injection} for short. The dCS gravity BH shadow will be the \textit{model} we fit to the Kerr injection. In order to determine the detectability of any GR deviation, we will first carry out a systematic bias analysis, and then a likelihood analysis.

\subsubsection{Systematic Bias Analysis} In this analysis, we fit for the spin parameter of the model with fixed values of $\zeta_{\dCS}$ and $\iota$ to search for a systematic bias in (a difference between the injected and the best-fit values of) the recovered spin and a deterioration in the goodness of fit. This is achieved by minimizing the relative $\chi^{2}$ between the injection and the model over the model spin parameter. To be conservative, we maximize the GR deviation of the model by fixing $\iota$ at the injected value and $\zeta_{\dCS}=0.5$, the maximum value allowed by the small-coupling approximation; this value also corresponds to a physical coupling constant of $\sqrt{|\alpha_{\dCS}|}\approx2\times10^{11}$cm using the Sagittarius A* mass $M=4.3\times10^{6}M_{\odot}$, which is a couple orders of magnitude smaller than the current upper bound $\sqrt{|\alpha_{\dCS}|}<10^{13}$cm~\cite{alihaimoud, PhysRevD.86.044037}. 

The model parameters have now been reduced to only the dimensionless spin $\chi$, and thus, the reduced $\chi^{2}$ is defined by
\begin{equation}
\chi^{2}_{\red}=\frac{\chi^{2}}{3}=\frac{1}{3}\sum_{i=1}^{3}\left(\frac{\alpha^{i}_{\dCS}\left(\chi\right)-\alpha^{i}_{\K}\left(\chi^{*}\right)}{\sigma_{i}}\right)^{2},
\end{equation}
where the BH shadow observables are $\alpha^{i}=[D,\langle R\rangle,A]$, $\alpha^{i}_{\K}$ is the injection, which depends only on the injected spin $\chi^{*}$, and $\alpha^{i}_{\dCS}$ is the model, which depends only on the model parameter $\chi$. The best-fit value of $\chi$ is that which minimizes the reduced $\chi^{2}$ for a given injected spin $\chi^{*}$. For the standard deviations, we make the choice $\left(\sigma_{D},\sigma_{\langle R\rangle},\sigma_{A}\right)=(0.05,0.005,0.002)$, which correspond to $\sim 10\%$ of the range of each observable in Fig.~\ref{fig:params}. This is somewhat of an arbitrary choice because there are currently no completed BH shadow observations, and thus, it is not obvious what the observational error will be given the capabilities of current observing campaigns.

\begin{figure}[htb]
\centering
\includegraphics[clip=true,width=\textwidth]{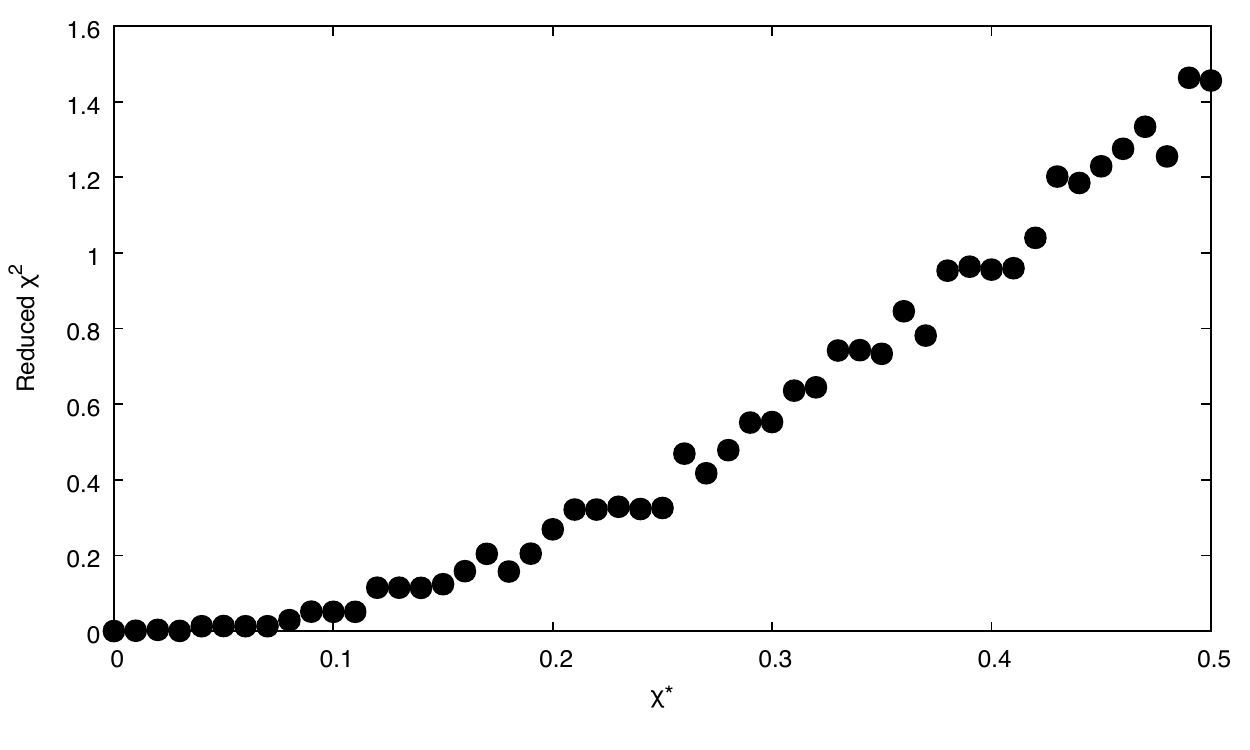}
\caption{\label{fig:chi2} Reduced $\chi^{2}$ of the best-fit dCS shadow as a function of injected spin $\chi^{*}$, for a fixed inclination angle $\iota=\pi/2=\iota^{*}$ and dCS coupling parameter $\zeta_{\dCS} = 0.5$. Observe that the reduced $\chi^{2}$ is smaller than unity for almost all values of the injected parameter, with the fit deteriorating as $\chi^{*} \gtrsim 0.45$}
\end{figure}
The $\chi^{2}_{\red}$ of the best-fit dCS shadow as a function of injected spin, for a fixed $\iota=\pi/2 = \iota^{*}$ is shown in Fig.~\ref{fig:chi2}. Observe that $\chi^{2}_{\red}$ is less than unity for almost all values of the injected spin. This means that the dCS BH shadow model can capture well the Kerr shadow injection. Moreover, the injected spin, not shown in the figure, is well recovered in all cases,~i.e.~the best-fit model's spin parameter is the same as the injected spin parameter to within a difference of $\delta\chi\approx0.01$. Thus, the deformation introduced into the BH shadow by dCS gravity is not significant enough to be measured by BH shadow observations. 
 
Observe, nonetheless, that the fit does become worse as the injected spin increases,~i.e.~the reduced $\chi^{2}$ becomes larger with increasing $\chi^{*}$. This trend is more significant for an inclination angle of $\iota=\pi/2=\iota^{*}$, which is to be expected as the effect of spin, and in turn the deformation due to dCS gravity, on the shape of the shadow is most significant at this inclination angle, vanishing when $\iota=0=\iota^{*}$.  This suggests the possibility that the deviation in the shadow introduced by dCS gravity of a rapidly-rotating BH may be large enough to be observable, and if so, a constraint could be placed in that scenario. Such a further study, though, requires a dCS BH solution that is exact or is found in the rapidly-rotating limit, such as that studied in~\cite{McNees:2015srl}, and this is beyond the scope of this work.

\subsubsection{Likelihood Analysis} In this analysis, we carry out a likelihood analysis to calculate the posterior probability distribution of the dimensionless coupling parameter $\zeta_{\dCS}$, marginalized over the other parameters (the spin and the inclination angle). That is, we calculate
\begin{equation}
P(\zeta)=\int \int \mathcal{L}(\chi,\iota,\zeta) \; p(\iota) \; p(\chi) \; d\iota \; d\chi,
\label{eq:posterior}
\end{equation}
where $\mathcal{L}(\chi,\iota,\zeta)$ is the likelihood function given by
\begin{equation}
\mathcal{L}(\chi,\iota,\zeta)\propto e^{-\chi^{2}(\chi,\iota,\zeta)/2}\,,
\end{equation}
and $p(\chi)$ and $p(\iota)$ are the prior distributions of the spin and the inclination angle. We recall that we here use flat priors in the range discussed in Sec.~\ref{subsec:characterization}. To calculate $\chi^{2}(\chi,\iota,\zeta_{\dCS})$, we use the same values for the standard deviations as in the systematic bias, $\left(\sigma_{D},\sigma_{\langle R\rangle},\sigma_{A}\right)=(0.05,0.005,0.002)$.

\begin{figure}[htb]
\centering
\includegraphics[]{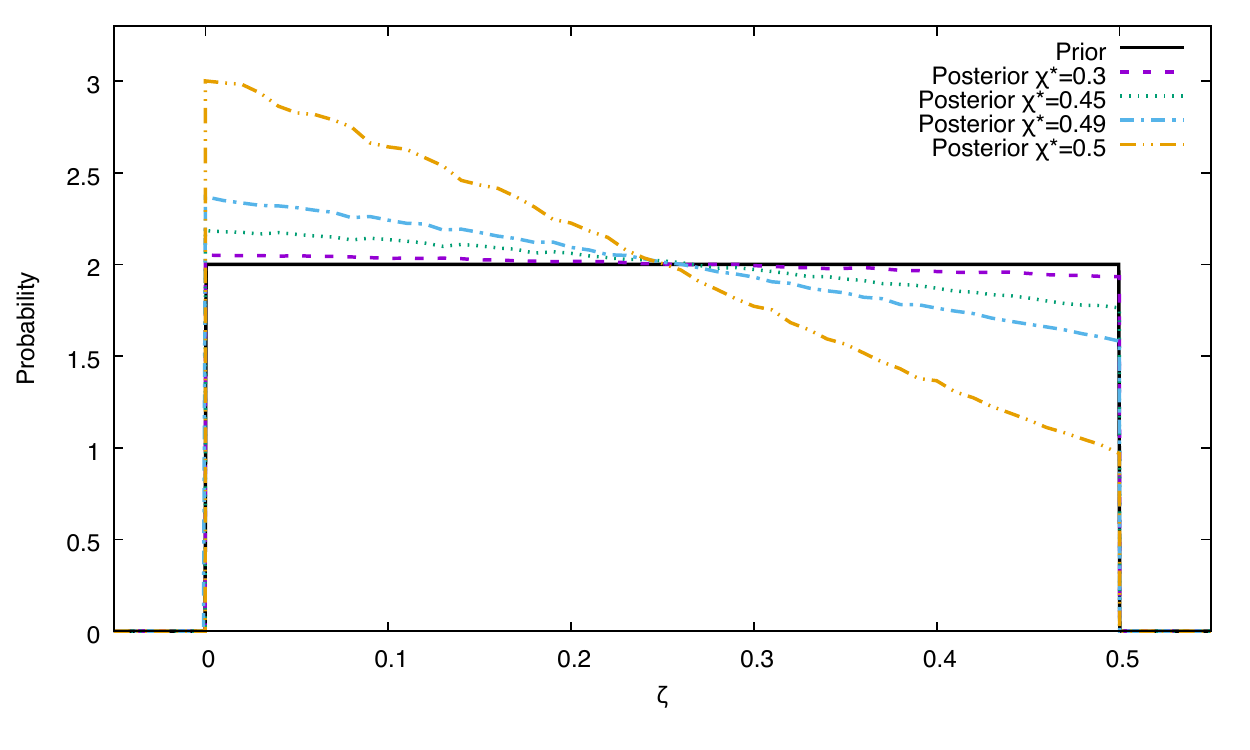}
\caption{\label{fig:post} Normalized prior (solid black line) and marginalized posteriors as a function of the modeled dimensionless deformation parameter $\zeta_{\dCS}$ marginalized over the modeled spin $\chi$ and inclination angle $\iota$ parameters for an injected inclination angle of $\iota=\pi/2$ and injected spins of $\chi^{*}=(0.3,0.45,0.49,0.5)$.}
\end{figure}

Figure~\ref{fig:post} shows the prior and the marginalized posterior of $\zeta_{\dCS}$ for injected spins of $\chi^{*} = (0.3,0.45,0.49,0.5)$ and injected inclination angle $\iota^{*} = \pi/2$. This figure re-enforces the systematic bias results: if the injected spins are small enough, the injection is not sufficiently informative to allow for constraints on $\zeta_{\dCS}$ because its marginalized posterior is almost identical to its prior. As we increase the injected spin, however, the marginalized posterior of $\zeta_{\dCS}$ becomes more and more peaked around zero, suggesting a constraint would then be possible. In particular, as the injected spin and the inclination angle increase, the marginalized posterior deviates from the flat prior, becoming significantly peaked for spins above $\chi^{*}\approx0.45$ and inclination angles above $\iota^{*}\approx5\pi/12=75^{\circ}$. Further work is needed to verify these conclusions using a numerical BH metric that is valid for moderate spins or a rapidly-rotating approximation~\cite{McNees:2015srl}.

\section{Discussion}
\label{sec:disc}

We have studied whether it is possible to place constraints on coupling constants in modified gravity theories using BH shadow observations. We focused on two theories within the broader class of quadratic gravity theories known as EdGB gravity and dCS gravity. Both are well-motivated modified gravity theories and have been studied extensively. We first argued that SMBH shadow observations cannot be used to place better-than-current constraints on EdGB gravity, as the strength of the current constraint and the large masses of SMBHs, the targets of BH shadow observing campaigns, would lead to a deformation from the Kerr solution that is much too small to be observed, now or in the future. DCS gravity, on the other hand, is not so well constrained and we have studied the BH shadows produced in this theory.

The BHs in dCS gravity were modeled using an approximate solution that is of fifth-order in a small spin expansion and linear in the coupling. We performed a resummation of the solution to remove unphysical behavior and other spurious features caused by the approximate nature of the solution. After simulating the BH shadows of this resummed solution, we analyzed whether the deviations of dCS shadows from Kerr shadows are large enough to be detectable by current BH shadow observing campaigns. We first studied whether the best-fit spin would be biased by extracting a Kerr BH observation with a dCS BH shadow model. We found that no bias was present unless the spin of the Kerr BH observation was moderately large. We then carried out a likelihood analysis to determine whether a Kerr shadow observation can update a flat prior on the dCS deformation parameter. We again found that the posterior on this deformation parameter is almost identical to the prior, unless the injected spin of the Kerr shadow is moderately large. We thus concluded that better-than-current constraints cannot be placed on dCS gravity with shadow observations of BHs with low spin. 

Our results also suggest that observations of BHs with large spin may lead to better-than-current constraints. This statement, however, requires a re-analysis of our results with a dCS BH metric that is valid outside the slow-rotation approximation, perhaps through a numerical dCS BH solution, or with a dCS BH solution derived in the rapidly-rotating limit~\cite{McNees:2015srl}. Another possible extension of our work is to lift the assumption of an ideal BH shadow,~i.e.~the photons that illuminate the region outside the shadow originate isotropically from spatial infinity. In reality, the majority of the illuminating photons originate from an accretion disk around the BH. Thus, the photons do not impinge on the BH isotropically and the presence of the disk itself may influence the ability to observe the shadow. One could begin to lift this assumption through GR magnetohydrodynamics simulations of the accretion disk~\cite{Medeiros:2016put} but around a dCS BH, and the reconstruction of the BH shadow in this more realistic analysis.

\ack

This work was supported by the NSF CAREER Grant PHY-1250636. DA also acknowledges support from the National Natural Science Foundation of China (NSFC), Grant No. U1531117, and Fudan University, Grant No. IDH1512060. NY also acknowledges support from NASA grants NNX16AB98G and 80NSSC17M0041. We would also like to acknowledge the support of the Research Group at Montana State University through their High Performance Computer Cluster \emph{Hyalite}.

\appendix

\section{BH Solution in dCS gravity}
\label{app:dCSsols}

We here provide the resummed dCS gravity BH solution in Boyer-Lindquist coordinates $(t,r,\theta,\phi)$ used throughout this work.

{\setlength{\mathindent}{0pt}
\begin{eqnarray}
g_{tt}^{\dCS}=&-\left(1-\frac{2Mr}{\Sigma_{\K}}\right)
+
\frac{201}{1792}\zeta_{\dCS} a^{2}\frac{M}{r^{3}}\left[\left(1+\frac{M}{r}+\frac{4474}{4221}\frac{M^{2}}{r^{2}}-\frac{2060}{469}\frac{M^{3}}{r^{3}}\right.\right.
\nonumber \\
&\left.\left.+\frac{1500}{469}\frac{M^{4}}{r^{4}}-\frac{2140}{201}\frac{M^{5}}{r^{5}}+\frac{9256}{201}\frac{M^{6}}{r^{6}}-\frac{5376}{67}\frac{M^{7}}{r^{7}}\right)\left(3\cos^{2}\theta-1\right)-\frac{70}{603}\frac{M^{2}}{r^{2}}\left(1\right.\right.
\nonumber \\
&\left.\left.+100\frac{M}{r}+194\frac{M^{2}}{r^{2}}+\frac{2220}{7}\frac{M^{3}}{r^{3}}-\frac{1512}{5}\frac{M^{4}}{r^{4}}\right)\right] 
-
\frac{701429}{23708160}\zeta_{\dCS} a^{4}\frac{M}{r^{5}}\left[\left(1\right.\right.
\nonumber \\
&\left.\left.+\frac{1013451}{701429}\frac{M}{r}+\frac{1154835}{701429}\frac{M^{2}}{r^{2}}-\frac{3346744}{701429}\frac{M^{3}}{r^{3}}+\frac{3992148}{701429}\frac{M^{4}}{r^{4}}-\frac{9591516}{701429}\frac{M^{5}}{r^{5}}\right.\right.
\nonumber \\
&\left.\left.+\frac{94091244}{701429}\frac{M^{6}}{r^{6}}-\frac{103967604}{701429}\frac{M^{7}}{r^{7}}-\frac{41345640}{701429}\frac{M^{8}}{r^{8}}+\frac{109734912}{701429}\frac{M^{9}}{r^{9}}\right)\right.
\nonumber \\
&\left.\times\left(35\cos^{4}\theta-30\cos^{2}\theta+3\right)+\frac{763980}{701429}\frac{r^{2}}{M^{2}}\left(1+\frac{M}{r}+\frac{51806}{12733}\frac{M^{2}}{r^{2}}+\frac{135383}{63665}\frac{M^{3}}{r^{3}}\right.\right.
\nonumber \\
&\left.\left.+\frac{309664}{38199}\frac{M^{4}}{r^{4}}-\frac{36264049}{381990}\frac{M^{5}}{r^{5}}-\frac{7873793}{38199}\frac{M^{6}}{r^{6}}-\frac{32533551}{63665}\frac{M^{7}}{r^{7}}+\frac{73025558}{63665}\frac{M^{8}}{r^{8}}\right.\right.
\nonumber \\
&\left.\left.-\frac{8708988}{12733}\frac{M^{9}}{r^{9}}-\frac{433800}{1819}\frac{M^{10}}{r^{10}}-\frac{3483648}{1819}\frac{M^{11}}{r^{11}}\right)\left(3\cos^{2}\theta-1\right)-\frac{61740}{701429}\left(1\right.\right.
\nonumber \\
&\left.\left.+\frac{2624}{35}\frac{M}{r}+\frac{492831}{3920}\frac{M^{2}}{r^{2}}+\frac{4430511}{980}\frac{M^{5}}{r^{5}}+\frac{330775}{168}\frac{M^{4}}{r^{4}}+\frac{1771487}{1680}\frac{M^{3}}{r^{3}}\right.\right.
\nonumber \\
&\left.\left.+\frac{15984}{5}\frac{M^{9}}{r^{9}}+\frac{667071}{70}\frac{M^{8}}{r^{8}}+\frac{6488861}{980}\frac{M^{7}}{r^{7}}-\frac{6957813}{980}\frac{M^{6}}{r^{6}}\right)\right],
\\
g_{rr}^{\dCS}=&\frac{\Sigma_{\K}}{\Delta_{\dCS}}
+
\frac{201}{1792}\zeta_{\dCS} a^{2}\frac{M}{r^{3}f_{\dCS}^{3}}\left[\left(1-\frac{953}{603}\frac{M}{r}-\frac{3968}{4221}\frac{M^{2}}{r^{2}}+\frac{115592}{4221}\frac{M^{3}}{r^{3}}-\frac{10420}{63}\frac{M^{4}}{r^{4}}\right.\right.
\nonumber \\
&\left.\left.+\frac{261116}{1407}\frac{M^{5}}{r^{5}}-\frac{61312}{201}\frac{M^{6}}{r^{6}}+\frac{393872}{201}\frac{M^{7}}{r^{7}}-\frac{161280}{67}\frac{M^{8}}{r^{8}}\right)\left(3\cos^{2}\theta-1\right)\right.
\nonumber \\
&\left.-\frac{350}{603}\frac{M}{r}\left(1+\frac{M}{r}+\frac{292}{5}\frac{M^{2}}{r^{2}}-\frac{446}{5}\frac{M^{3}}{r^{3}}-\frac{7584}{175}\frac{M^{4}}{r^{4}}-\frac{135024}{175}\frac{M^{5}}{r^{5}}\right.\right.
\nonumber \\
&\left.\left.+\frac{34992}{25}\frac{M^{6}}{r^{6}}\right)\right]
-
\frac{94699}{4741632}\zeta_{\dCS} a^{4}\frac{M}{r^{5}f_{\dCS}}\left[\left(1-\frac{916004}{473495}\frac{M}{r}+\frac{2411573}{473495}\frac{M^{2}}{r^{2}}\right.\right.
\nonumber \\
&\left.\left.+\frac{16109646}{473495}\frac{M^{3}}{r^{3}}+\frac{2585472}{43045}\frac{M^{4}}{r^{4}}-\frac{23898480}{94699}\frac{M^{5}}{r^{5}}-\frac{314374068}{473495}\frac{M^{6}}{r^{6}}\right.\right.
\nonumber \\
&\left.\left.-\frac{41960268}{43045}\frac{M^{7}}{r^{7}}+\frac{1261951488}{473495}\frac{M^{8}}{r^{8}}\right)\left(35\cos^{4}\theta-30\cos^{2}\theta+3\right)\right.
\nonumber \\
&\left.+\frac{152796}{94699}\frac{r^{2}}{M^{2}f_{\dCS}}\left(1+\frac{1084}{1819}\frac{M}{r}+\frac{62621}{12733}\frac{M^{2}}{r^{2}}+\frac{56726}{1819}\frac{M^{3}}{r^{3}}-\frac{2116475}{38199}\frac{M^{4}}{r^{4}}\right.\right.
\nonumber \\
&\left.\left.-\frac{112168723}{381990}\frac{M^{5}}{r^{5}}-\frac{36858343}{190995}\frac{M^{6}}{r^{6}}+\frac{3546621}{9095}\frac{M^{7}}{r^{7}}-\frac{47131846}{63665}\frac{M^{8}}{r^{8}}\right.\right.
\nonumber \\
&\left.\left.-\frac{5777844}{12733}\frac{M^{9}}{r^{9}}-\frac{32693976}{1819}\frac{M^{10}}{r^{10}}+\frac{80123904}{1819}\frac{M^{11}}{r^{11}}\right)\left(3\cos^{2}\theta-1\right)\right.
\nonumber \\
&\left.-\frac{61740}{94699}\frac{r}{Mf_{\dCS}^{2}}\left(1+\frac{577}{175}\frac{M}{r}+\frac{8113}{1200}\frac{M^{2}}{r^{2}}-\frac{109309}{11760}\frac{M^{3}}{r^{3}}+\frac{2125311}{3920}\frac{M^{4}}{r^{4}}\right.\right.
\nonumber \\
&\left.\left.+\frac{267403}{1470}\frac{M^{5}}{r^{5}}+\frac{2001821}{420}\frac{M^{6}}{r^{6}}-\frac{19927289}{980}\frac{M^{7}}{r^{7}}+\frac{22161021}{980}\frac{M^{8}}{r^{8}}-\frac{12553726}{245}\frac{M^{9}}{r^{9}}\right.\right.
\nonumber \\
&\left.\left.+\frac{249993}{5}\frac{M^{10}}{r^{10}}+\frac{735264}{25}\frac{M^{11}}{r^{11}}\right)\right]
+
\delta g_{rr},
\\
g_{\theta\theta}^{\dCS}=&\Sigma_{\K}
+
\frac{201}{1792}\zeta_{\dCS} a^{2}\frac{m}{r}\left(1+\frac{1420}{603}\frac{M}{r}+\frac{18908}{4221}\frac{M^{2}}{r^{2}}+\frac{1480}{603}\frac{M^{3}}{r^{3}}+\frac{22460}{1407}\frac{M^{4}}{r^{4}}\right.
\nonumber \\
&\left.+\frac{3848}{201}\frac{M^{5}}{r^{5}}+\frac{5376}{67}\frac{M^{6}}{r^{6}}\right)\left(3\cos^{2}-1\right)
-
\frac{94699}{4741632}\zeta_{\dCS} a^{4}\frac{M}{r^{3}}\left[\left(1+\frac{2984191}{1420485}\frac{M}{r}\right.\right.
\nonumber \\
&\left.\left.+\frac{2339824}{473495}\frac{M^{2}}{r^{2}}+\frac{94116}{8609}\frac{M^{3}}{r^{3}}+\frac{45539276}{1420485}\frac{M^{4}}{r^{4}}+\frac{16610916}{473495}\frac{M^{5}}{r^{5}}+\frac{7853220}{94699}\frac{M^{6}}{r^{6}}\right.\right.
\nonumber \\
&\left.\left.-\frac{31810968}{473495}\frac{M^{7}}{r^{7}}-\frac{109734912}{473495}\frac{M^{8}}{r^{8}}\right)\left(35\cos^{4}\theta-30\cos^{2}\theta+3\right)+\frac{152796}{94699}\frac{r^{2}}{M^{2}}\left(1\right.\right.
\nonumber \\
&\left.\left.+\frac{17455}{7276}\frac{M}{r}+\frac{148755}{25466}\frac{M^{2}}{r^{2}}+\frac{52999}{3638}\frac{M^{3}}{r^{3}}+\frac{3438929}{76398}\frac{M^{4}}{r^{4}}+\frac{5163387}{63665}\frac{M^{5}}{r^{5}}\right.\right.
\nonumber \\
&\left.\left.+\frac{14491811}{190995}\frac{M^{6}}{r^{6}}-\frac{5632}{85}\frac{M^{7}}{r^{7}}+\frac{6094488}{12733}\frac{M^{8}}{r^{8}}+\frac{1232136}{1819}\frac{M^{9}}{r^{9}}+\frac{3483648}{1819}\frac{M^{10}}{r^{10}}\right)\right.
\nonumber \\
&\left.\times\left(3\cos^{2}\theta-1\right)-\frac{709128}{473495}\left(1+\frac{104533}{19296}\frac{M}{r}+\frac{583357}{45024}\frac{M^{2}}{r^{2}}+\frac{311763}{7504}\frac{M^{3}}{r^{3}}\right.\right.
\nonumber \\
&\left.\left.+\frac{3112171}{33768}\frac{M^{4}}{r^{4}}+\frac{24899}{1608}\frac{M^{5}}{r^{5}}-\frac{2538845}{11256}\frac{M^{6}}{r^{6}}-\frac{190101}{268}\frac{M^{7}}{r^{7}}-\frac{18648}{67}\frac{M^{8}}{r^{8}}\right)\right],
\\
g_{\phi\phi}^{\dCS}=&\left(r^{2}+a^{2}+\frac{2Ma^{2}r\sin^{2}\theta}{\Sigma_{\K}}\right)\sin^{2}\theta
+
\frac{201}{1792}\zeta_{\dCS} a^{2}\frac{M}{r}\sin^{2}\theta\left(1+\frac{1420}{603}\frac{M}{r}\right.
\nonumber \\
&\left.+\frac{18908}{4221}\frac{M^{2}}{r^{2}}+\frac{1480}{603}\frac{M^{3}}{r^{3}}+\frac{22460}{1407}\frac{M^{4}}{r^{4}}+\frac{3848}{201}\frac{M^{5}}{r^{5}}+\frac{5376}{67}\frac{M^{6}}{r^{6}}\right)\left(3\cos^{2}\theta-1\right)
\nonumber \\
&-
\frac{701429}{23708160}\zeta_{\dCS} a^{4}\frac{M}{r^{3}}\sin^{2}\theta\left[\left(1+\frac{5962075}{2104287}\frac{M}{r}+\frac{4434376}{701429}\frac{M^{2}}{r^{2}}+\frac{7777884}{701429}\frac{M^{3}}{r^{3}}\right.\right.
\nonumber \\
&\left.\left.+\frac{59811476}{2104287}\frac{M^{4}}{r^{4}}+\frac{28251588}{701429}\frac{M^{5}}{r^{5}}+\frac{66282516}{701429}\frac{M^{6}}{r^{6}}+\frac{4767336}{701429}\frac{M^{7}}{r^{7}}\right.\right.
\nonumber \\
&\left.\left.-\frac{109734912}{701429}\frac{M^{8}}{r^{8}}\right)\left(35\cos^{4}\theta-30\cos^{2}\theta+3\right)+\frac{763980}{701429}\frac{r^{2}}{M^{2}}\left(1+\frac{17455}{7276}\frac{M}{r}\right.\right.
\nonumber \\
&\left.\left.+\frac{106545}{25466}\frac{M^{2}}{r^{2}}+\frac{26739}{3638}\frac{M^{3}}{r^{3}}+\frac{2275289}{76398}\frac{M^{4}}{r^{4}}+\frac{3958987}{63665}\frac{M^{5}}{r^{5}}+\frac{463783}{11235}\frac{M^{6}}{r^{6}}\right.\right.
\nonumber \\
&\left.\left.-\frac{9607568}{63665}\frac{M^{7}}{r^{7}}+\frac{3592968}{12733}\frac{M^{8}}{r^{8}}+\frac{748296}{1819}\frac{M^{9}}{r^{9}}+\frac{3483648}{1819}\frac{M^{10}}{r^{10}}\right)\left(3\cos^{2}\theta-1\right)\right.
\nonumber \\
&\left.-\frac{3013647}{2805716}\frac{m}{r}\left(1+\frac{508855}{143507}\frac{M}{r}+\frac{4070102}{143507}\frac{M^{2}}{r^{2}}+\frac{9629484}{143507}\frac{M^{3}}{r^{3}}-\frac{4806660}{143507}\frac{M^{4}}{r^{4}}\right.\right.
\nonumber \\
&\left.\left.-\frac{46475868}{143507}\frac{M^{5}}{r^{5}}-\frac{16783848}{20501}\frac{M^{6}}{r^{6}}-\frac{5370624}{20501}\frac{M^{7}}{r^{7}}\right)\right],
\\
g_{t\phi}^{\dCS}=&-\frac{2Mar\sin^{2}\theta}{\Sigma_{\K}}
+
\frac{5}{8}\zeta_{\dCS} a\frac{M^{4}}{r^{4}}\sin^{2}\theta\left(1+\frac{12}{7}\frac{M}{r}+\frac{27}{10}\frac{M^{2}}{r^{2}}\right)
\nonumber \\
&-
\frac{8819}{56448}\zeta_{\dCS} a^{3}\frac{M}{r^{3}}\sin^{2}\theta\left[\left(1+\frac{24875}{35276}\frac{M}{r}+\frac{95}{17638}\frac{M^{2}}{r^{2}}+\frac{90188}{26457}\frac{M^{3}}{r^{3}}+\frac{684818}{26457}\frac{M^{4}}{r^{4}}\right.\right.
\nonumber \\
&\left.\left.+\frac{385542}{8819}\frac{M^{5}}{r^{5}}+\frac{418572}{8819}\frac{M^{6}}{r^{6}}-\frac{508032}{8819}\frac{M^{7}}{r^{7}}\right)\left(3\cos^{2}\theta-1\right)+\frac{2}{5}\left(1+\frac{60155}{35276}\frac{M}{r}\right.\right.
\nonumber \\
&\left.\left.+\frac{8545}{8819}\frac{M^{2}}{r^{2}}-\frac{19828}{26457}\frac{M^{3}}{r^{3}}+\frac{563669}{26457}\frac{M^{4}}{r^{4}}+\frac{549630}{8819}\frac{M^{5}}{r^{5}}+\frac{873180}{8819}\frac{M^{6}}{r^{6}}\right)\right]
\nonumber \\
&+
\frac{65029949}{1738598400}\zeta_{\dCS} a^{5}\frac{M}{r^{5}}\sin^{2}\theta\left[\left(1+\frac{247489546}{195089847}\frac{M}{r}-\frac{192857740}{585269541}\frac{M^{2}}{r^{2}}\right.\right.
\nonumber \\
&\left.\left.+\frac{201416960}{195089847}\frac{M^{3}}{r^{3}}+\frac{6952033840}{195089847}\frac{M^{4}}{r^{4}}+\frac{49673623120}{585269541}\frac{M^{5}}{r^{5}}+\frac{8477276720}{65029949}\frac{M^{6}}{r^{6}}\right.\right.
\nonumber \\
&\left.\left.-\frac{7984872720}{65029949}\frac{M^{7}}{r^{7}}-\frac{5714422560}{65029949}\frac{M^{8}}{r^{8}}+\frac{8047226880}{65029949}\frac{M^{9}}{r^{9}}\right)\right.
\nonumber \\
&\left.\times\left(35\cos^{4}\theta-30\cos^{2}\theta+3\right)+\frac{1056250525}{585269541}\frac{r^{2}}{M^{2}}\left(1+\frac{10036795}{7681822}\frac{M}{r}\right.\right.
\nonumber \\
&\left.\left.+\frac{949643961}{211250105}\frac{M^{2}}{r^{2}}+\frac{1196741284}{211250105}\frac{M^{3}}{r^{3}}+\frac{304195064}{42250021}\frac{M^{4}}{r^{4}}+\frac{646950168}{211250105}\frac{M^{5}}{r^{5}}\right.\right.
\nonumber \\
&\left.\left.+\frac{19300145456}{211250105}\frac{M^{6}}{r^{6}}+\frac{1091987984}{3840911}\frac{M^{7}}{r^{7}}+\frac{13626382752}{19204555}\frac{M^{8}}{r^{8}}-\frac{914366016}{3840911}\frac{M^{9}}{r^{9}}\right.\right.
\nonumber \\
&\left.\left.-\frac{290957184}{3840911}\frac{M^{10}}{r^{10}}-\frac{3511517184}{3840911}\frac{M^{11}}{r^{11}}\right)\left(3\cos^{2}\theta-1\right)+\frac{422500210}{585269541}\frac{r^{2}}{M^{2}}\left(1\right.\right.
\nonumber \\
&\left.\left.+\frac{3368875}{7681822}\frac{M}{r}+\frac{539981961}{211250105}\frac{M^{2}}{r^{2}}+\frac{63963088}{211250105}\frac{M^{3}}{r^{3}}-\frac{28203665}{84500042}\frac{M^{4}}{r^{4}}\right.\right.
\nonumber \\
&\left.\left.-\frac{218979789}{84500042}\frac{M^{5}}{r^{5}}+\frac{6554146711}{42250021}\frac{M^{6}}{r^{6}}+\frac{1870270010}{3840911}\frac{M^{7}}{r^{7}}+\frac{3798260802}{3840911}\frac{M^{8}}{r^{8}}\right.\right.
\nonumber \\
&\left.\left.-\frac{1514962386}{3840911}\frac{M^{9}}{r^{9}}-\frac{2505947220}{3840911}\frac{M^{10}}{r^{10}}-\frac{1184222592}{3840911}\frac{M^{11}}{r^{11}}\right)\right].
\end{eqnarray}
}
Here $\Sigma_{\K}\equiv r^2+a^2\cos^2\theta$, $f_{\dCS}=1-r^{\dCS}_{\hor}/r$ where
\begin{equation}
r^{\dCS}_{\hor}=2M\left(1-\frac{1}{4}\chi^{2}-\frac{1}{16}\chi^{4}\right)-\frac{915}{28672}\zeta_{\dCS} M\chi^{2}\left(1+\frac{351479}{439200}\chi^{2}\right)
\end{equation}
is the event horizon radius of this dCS gravity BH solution when all calculations are expanded in a slowly-rotating and small-coupling approximation and truncated to $\mathcal{O}\left(\zeta_{\dCS},\chi^{5}\right)$,
\begin{equation}
\Delta_{\dCS}=\Delta_{\K}-\frac{915}{14336}\zeta_{\dCS} a^{2}\left(1+\frac{351479}{439200}\frac{a^{2}}{M^{2}}\right),
\end{equation}
where $\Delta_{\K}\equiv r^{2}-2Mr+a^{2}$, and
{\setlength{\mathindent}{0pt}
\begin{eqnarray}
\delta g_{rr}&=\frac{915}{14336}\zeta_{\dCS}\chi^{2}\frac{M^{2}}{r^{2}f^{2}_{\dCS}}+\frac{2717}{14336}\zeta_{\dCS}\chi^{4}\frac{M^{4}}{r^{4}f^{3}_{\dCS}}\left[\left(1+\frac{402}{2717}\frac{M}{r}-\frac{1704}{19019}\frac{M^{2}}{r^{2}}+\frac{458960}{19019}\frac{M^{3}}{r^{3}}\right.\right.
\nonumber \\
&\left.\left.-\frac{1874640}{19019}\frac{M^{4}}{r^{4}}-\frac{6768}{209}\frac{M^{5}}{r^{5}}-\frac{911712}{2717}\frac{M^{6}}{r^{6}}+\frac{2903040}{2717}\frac{M^{7}}{r^{7}}\right)\left(3\cos^{2}\theta-1\right)\right.
\nonumber \\
&\left.+\frac{631}{6240}\frac{r^{2}}{M^{2}}\left(1+\frac{175442}{131879}\frac{M}{r}-\frac{732000}{131879}\frac{M^{2}}{r^{2}}-\frac{785600}{131879}\frac{M^{3}}{r^{3}}-\frac{75008}{6941}\frac{M^{4}}{r^{4}}\right.\right.
\nonumber \\
&\left.\left.-\frac{126195776}{395637}\frac{M^{5}}{r^{5}}-\frac{593814080}{2769459}\frac{M^{6}}{r^{6}}+\frac{511400640}{923153}\frac{M^{7}}{r^{7}}+\frac{10560693760}{2769459}\frac{M^{8}}{r^{8}}\right.\right.
\nonumber \\
&\left.\left.+\frac{134859520}{20823}\frac{M^{9}}{r^{9}}-\frac{25506929920}{923153}\frac{M^{10}}{r^{10}}+\frac{28366106880}{923153}\frac{M^{11}}{r^{11}}-\frac{64275077120}{923153}\frac{M^{12}}{r^{12}}\right.\right.
\nonumber \\
&\left.\left.+\frac{8959749120}{131879}\frac{M^{13}}{r^{13}}+\frac{5270372352}{131879}\frac{M^{14}}{r^{14}}\right)\right].
\end{eqnarray}
}
%

\section{Black Hole Shadow in the Kerr Spacetime}
\label{app:shadow}

We here show how to calculate the photon sphere and, in turn, the BH shadow in the Kerr spacetime in Boyer-Lindquist coordinates.

We begin with the Hamilton-Jacobi equation
\begin{equation}
\frac{\partial S}{\partial \lambda}=\frac{1}{2}g^{ab}\frac{\partial S}{\partial x^a}\frac{\partial S}{\partial x^b},\label{HJE}
\end{equation}
where $S$ is the Jacobi action, $\lambda$ is the affine parameter, and $x^a$ are generalized coordinates. If we assume separability and note that we only care about null geodesics, the Jacobi action can be written as
\begin{equation}
S=-Et+L_z\phi+S_r(r)+S_\theta(\theta).
\end{equation}
Inserting this ansatz into Eq.~(\ref{HJE}), we find the partial differential equation
\begin{eqnarray}
2\frac{\partial S}{\partial\lambda}=0=&g^{tt}E^2-2g^{t\phi}EL_z+g^{\phi\phi}L_z^2
\nn \\
&+g^{rr}\left(\frac{dS_r}{dr}\right)^2+g^{\theta\theta}\left(\frac{dS_\theta}{d\theta}\right)^2\,,
\end{eqnarray}
which through separation of variables becomes
\begin{eqnarray}
\Delta_{\K}\left(\frac{dS_r}{dr}\right)^2=\frac{1}{\Delta_{\K}}\left[E\left(r^2+a^2\right)-aL_z\right]^2-\left(L_z-aE\right)^2-\mathcal{Q}&,\label{dSr}
\\
\left(\frac{dS_\theta}{d\theta}\right)^2=\mathcal{Q}+\cos^2\theta\left[a^2E^2-\frac{L_z^2}{\sin^2\theta}\right],&\label{dSth}
\end{eqnarray}
where $\mathcal{Q}$ is the Carter constant.

The null-geodesic equations for the $r(\lambda)$ and $\theta(\lambda)$ components of the null trajectories can be found by noting that $dS/dr=p_r=g_{rr} (dr/d\lambda)$ and $dS/d\theta=p_\theta=g_{\theta\theta} (d\theta/d\lambda)$. The equations are then simply
\begin{eqnarray}
\Sigma_{\K}\frac{dr}{d\lambda}&=\pm\sqrt{\mathcal{R}}, \label{eq:drdlambda}
\\
\Sigma_{\K}\frac{d\theta}{d\lambda}&=\pm\sqrt{\Theta}\,, \label{eq:dthetadlambda}
\end{eqnarray}
where we have defined the two functions
\begin{eqnarray}
\mathcal{R}(r)&\equiv\left[E\left(r^2+a^2\right)-aL_z\right]^2-\Delta_{\K}\left[\mathcal{Q}+\left(L_z-aE\right)^2\right],\label{eqR}
\\
\Theta(\theta)&\equiv\mathcal{Q}+\cos^2\theta\left[a^2E^2-\frac{L_z^2}{\sin^2\theta}\right].
\label{eq:Theta}
\end{eqnarray}

Unstable spherical photon orbits are defined by the conditions
\begin{eqnarray}
\mathcal{R}=0, \quad
\frac{d\mathcal{R}}{dr}=0,  
\label{eq:spherical-photon-orbit0}
\\
\Theta\geq0.
\label{eq:spherical-photon-orbit}
\end{eqnarray}
For simplicity, we define the conserved quantities $\xi\equiv L_z/E$ and $\eta\equiv Q/E^2$, and solve for each using Eq.~(\ref{eqR}) and Eq.~\eref{eq:spherical-photon-orbit0} to find
\begin{eqnarray}
\xi_{\sph}&=\frac{r_{\sph}^2+a^2}{a}-\frac{2\Delta_{\K} r_{\sph}}{a\left(r_{\sph}-m\right)},\label{xisol}
\\
\eta_{\sph}&=-\frac{r_{\sph}^3\left[r_{\sph}\left(r_{\sph}-3m\right)^2-4a^2m\right]}{a^2\left(r_{\sph}-m\right)^2},\label{etasol}
\end{eqnarray}
where $r_{\sph}$ is the constant radius of the unstable spherical orbits. This radius is  constrained by the condition in Eq.~\eref{eq:spherical-photon-orbit}, which we can simplify by rewriting $\Theta$ in terms of $\xi$ and $\eta$ as
\begin{eqnarray}
\frac{\Theta}{E^2}&=\mathcal{J}-\left(a\sin\theta-\xi\csc\theta\right)^2,
\end{eqnarray}
where
\begin{equation}
\mathcal{J}=\eta+\left(a-\xi\right)^2.
\end{equation}
Therefore, the $\Theta\geq0$ condition for unstable spherical orbits implies the necessary (but not sufficient) condition $\mathcal{J}\geq0$. Substituting Eqs.~(\ref{xisol}) and~(\ref{etasol}) into the above gives the condition
\begin{equation}
\mathcal{J}=\frac{4r_{\sph}^2\Delta_{\K}}{\left(r_{\sph}-m\right)^2} \geq 0\,,
\end{equation}
which reduces simply to $\Delta_{\K}\geq0$ or $r_{\sph}\geq m+\sqrt{m^2-a^2}$, which is the Kerr horizon radius.

The BH shadow boundary is defined as the sky projection of the photon sphere as observed at spatial infinity. The conserved parameters $\xi$ and $\eta$ can be related to the celestial coordinates of the observer at infinity via
\begin{eqnarray}
\alpha&=\lim_{r\rightarrow\infty}\frac{-rp^{(\phi)}}{p^{(t)}}=-\frac{\xi_{\sph}}{\sin \iota},
\\
\beta&=\lim_{r\rightarrow\infty}\frac{rp^{(\theta)}}{p^{(t)}}=\left(\eta_{\sph}+a^2\cos^2 \iota-\xi_{\sph}^2\cot^2 \iota\right)^{1/2}.
\end{eqnarray} 

The separatrix between ingoing and outgoing photon geodesics, what we call the boundary of the BH shadow, can then be constructed by plotting $(\alpha,\beta)$ by varying $r_{\sph}$ from $r = m + \sqrt{m^{2} - a^{2}}$ to $r=4.5 m$; the former is the Kerr horizon radius and the latter is the largest radius for which closed photon orbits are possible for all values of spin.

\section*{References}
\bibliography{biblio}
\bibliographystyle{iopart-num}

\end{document}